\documentclass[aps, pre,twocolumn,superscriptaddress, 10pt]{revtex4-2}

\usepackage{graphicx}
\usepackage{bm}
\usepackage[separate-uncertainty = true,multi-part-units=single]{siunitx}
\usepackage{xspace}
\usepackage{amssymb,amsfonts,amsmath}
\usepackage{siunitx}
\usepackage{hyperref}
\usepackage{comment}
\usepackage{bm}
\usepackage{mathtools}
\usepackage{xcolor}

\hypersetup{
    colorlinks=true,
    linkcolor=blue,
    filecolor=magenta,      
    urlcolor=cyan,
    citecolor = magenta
    }

\usepackage[mathlines]{lineno}
\modulolinenumbers[10]
\newcommand*\patchAmsMathEnvironmentForLineno[1]{
  \expandafter\let\csname old#1\expandafter\endcsname\csname #1\endcsname
  \expandafter\let\csname oldend#1\expandafter\endcsname\csname end#1\endcsname
  \renewenvironment{#1}
     {\linenomath\csname old#1\endcsname}
     {\csname oldend#1\endcsname\endlinenomath}}
\newcommand*\patchBothAmsMathEnvironmentsForLineno[1]{
  \patchAmsMathEnvironmentForLineno{#1}
  \patchAmsMathEnvironmentForLineno{#1*}}
\AtBeginDocument{
\patchBothAmsMathEnvironmentsForLineno{equation}
\patchBothAmsMathEnvironmentsForLineno{align}
\patchBothAmsMathEnvironmentsForLineno{flalign}
\patchBothAmsMathEnvironmentsForLineno{alignat}
\patchBothAmsMathEnvironmentsForLineno{gather}
\patchBothAmsMathEnvironmentsForLineno{multline}
}

\linenumbers

\begin{document}

\title{Acceleration of enzymatic reaction-diffusion kinetics by intermediate state}

\newcommand{\apph}{Department of Applied Physics, Graduate School of Engineering, Tohoku University, 980-8579 Sendai, Japan}

\author{Akihiro Fukuda}
\affiliation{\apph}
\author{Yohei Nakayama}
\affiliation{\apph}
\author{Shoichi Toyabe}
\email{toyabe@tohoku.ac.jp}
\affiliation{\apph}
\date{\today}

\newcommand{\Uo}{U_{\mathrm{p}, n}}
\newcommand{\ko}{k_\mathrm{p}}
\newcommand{\Uint}{U_{\mathrm{i}, n}}
\newcommand{\kint}{k_\mathrm{i}}
\newcommand{\xint}{x_\mathrm{i}}
\newcommand{\Gtot}{G}
\newcommand{\Gint}{G_\mathrm{i}}

\newcommand{\woip}{w_{\mathrm{p}, n}^{+}}
\newcommand{\wiop}{w_{\mathrm{i}, n}^{+}}
\newcommand{\woim}{w_{\mathrm{p}, n+1}^{-}}
\newcommand{\wiom}{w_{\mathrm{i}, n}^{-}}
\newcommand{\woipeff}{w_\mathrm{p}^{\mathrm{+, eff}}}
\newcommand{\wiomeff}{w_\mathrm{i}^{\mathrm{-, eff}}}
\newcommand{\wiopeff}{w_\mathrm{i}^{\mathrm{+, eff}}}
\newcommand{\woimeff}{w_\mathrm{p}^{\mathrm{-, eff}}}

\newcommand{\wzero}{\tilde w}
\newcommand{\kB}{k_\mathrm{B}}
\newcommand{\kBT}{k_\mathrm{B}T}
\newcommand{\VPMF}{V_\mathrm{PMF}}
\newcommand{\Jmax}{J^*}
\newcommand{\aopt}{\alpha^*}
\newcommand{\etamax}{\eta^*}

\newcommand{\Pbss}{P_{\mathrm{p}, n}^{\mathrm{ss}}}
\newcommand{\Pintss}{P_{\mathrm{i} ,n}^{\mathrm{ss}}}

\newcommand{\Pb}{P_{\mathrm{p}, n}}
\newcommand{\Pint}{P_{\mathrm{i}, n}}

\newcommand{\Wext}{W_{\mathrm{ext}}}

\begin{abstract}
Biological molecular motors are high-performance nanomachines that convert chemical energy into mechanical motion via chemomechanical coupling.
Their reaction cycles typically comprise a series of intermediate chemical states between the initial and final primary states.
However, the influence of these intermediate states on motor performance has not yet been fully explored.
In this study, we investigate the impact of intermediate states on the motor kinetics using a reaction-diffusion model.
In most cases, the intermediate states accelerate the motor by lowering the effective barrier height.
This acceleration is particularly pronounced when an external load is applied to the motor, implying the practical importance of the intermediate states.
The intermediate states can also slow down the reaction in some cases, such as the slow reaction limit with asymmetric kinetics.
Our findings provide practical insights into the design principles behind the high performance of biological molecular motors, as well as the development of efficient artificial molecular motors.
\end{abstract}

\maketitle

\section{Introduction}

Biological molecular motors generate unidirectional motions by coupling chemical reactions with mechanical motions~\cite{cell, Howard}.
The developments in single-molecule observation techniques have demonstrated that the mechanical motions of the motors are often stepwise, thus, the chemical states can be treated as discrete events~\cite{Svoboda1993, Funatsu1995, Kodera2010}.
Consider a motor that produces mechanical movement by hydrolysing adenosine triphosphate (ATP) molecules, such as kinesin, myosin, and F$_1$-ATPase (F$_1$).
The ATP hydrolysis cycle proceeds in multiple steps.
An ATP binds to the motor and is hydrolyzed to adenosine diphosphate (ADP) and inorganic phosphate (P$_\mathrm{i}$), which then dissociate from the motor. Consequently, multiple intermediate states connecting the initial and final primary states are often observed during a reaction cycle in single-molecule observation assays. 

The performance of the motor, including the speed, efficiency, and robustness, is expected to depend significantly on the relative positions of the intermediate states, i.e., the free energy difference and the spatial distance.
However, it is often not obvious how the positions of the intermediate states are chosen through evolution.
For example, F$_1$, an ATP-driven rotational molecular motor, drives 120$^\circ$ rotation per a single turnover of ATP hydrolysis~\cite{Abrahams1994Nature, Noji1997Nature, Yasuda1998, Noji_review}.
The 120$^\circ$ rotation is resolved into sub-rotation steps with the step sizes varying from species to species~\cite{Yasuda2001, ZarcoZavala2020, Watanabe2023}.
F$_1$ from thermophiles shows 80$^\circ$ and 40$^\circ$ substeps~\cite{Yasuda2001}, while human mitochondrial F$_1$ shows three substeps: 65$^\circ$, 25$^\circ$, and 30$^\circ$~\cite{hMF1}.
The fact that the substep sizes differ despite these motors having the same enzymatic functions is intriguing.

So far, only limited research has addressed how the intermediate states influence kinetics.
Previous studies based on Markov jump process have emphasized the importance of the barrier height between states, which is modulated by intermediate states~\cite{Brown2017, Wagoner2019}.
These studies focus on the general characterization of the effect of the intermediate states and do not consider diffusion in spatial degrees of freedom.
However, the time scales of the reactions and the diffusion in the potential, which involves large and slow conformational changes, are often comparable, leading to complex but rich nonequilibrium dynamics.
Therefore, diffusion is an essential factor in determining the kinetics of molecular motors, and its explicit implementation is necessary when discussing practical situations~\cite{Jlicher1997, Wang1998, Elston1998, Keller2000, Toyabe2012, KAWAGUCHI20142450}.
In addition, reaction-diffusion models can readily incorporate the modulation of kinetics by external load, while it is not straightforward to consider that in the Markov jump process that lacks spatial degree of freedom.

\begin{figure}[t]
    \centering
    \includegraphics{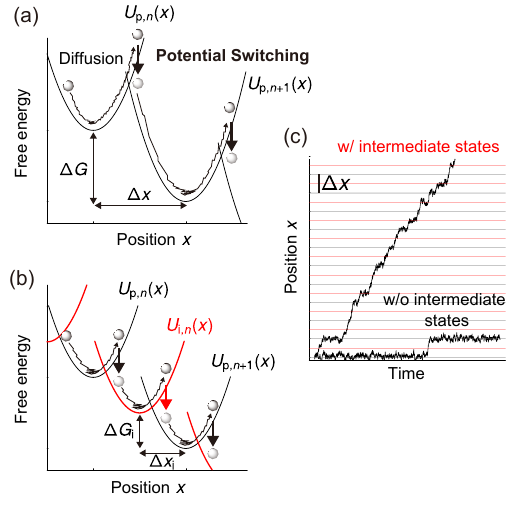}
    \caption{Reaction-diffusion model of molecular motors. (a) The free energy landscape including the motor and the solution of the fuel, such as ATP. The black potentials $\Uo(x)$ correspond to the primary states.
    A Brownian particle, which represents the motor, diffuses on the potential and jumps between different chemical states upon reactions.
    (b) The intermediate states $\Uint(x)$ (red) are located with a shift of $(\Delta \xint, \Delta \Gint)$ from $(n+1)$-th primary states.
    (c) An example of the trajectories obtained from the overdamped Langevin equation [Eq.~\eqref{eq:Langevin_eq}] with potential switching with or without the intermediate states.
    }
     \label{fig:intro}
\end{figure}

In this study, we investigate the effects of intermediate states on the kinetics of the molecular motors using a reaction-diffusion model~[Fig.~\ref{fig:intro}] to understand the design principle behind the high performance of biological molecular motors.
Specifically, we examine whether the intermediate states accelerate the motor speed or not, and also how they should be positioned to maximize speed.

\section{Methods}
\subsection{Reaction-diffusion model}
We consider a reaction-diffusion model to describe the motions of molecular motors with one spatial degree of freedom $x$ and one chemical degree of freedom for simplicity~\cite{Jlicher1997, Wang1998, Elston1998, Keller2000, Toyabe2012, KAWAGUCHI20142450}.
Particularly, we think of a system in which distinct chemical states are associated with different potential landscapes, and switching between neighboring potentials corresponds to chemical reactions.
A similar framework has also been used for general chemical reactions, where the potential is described by the function of a properly defined reaction coordinate~\cite{Marcus}.

\begin{figure}[b]
    \centering
    \includegraphics{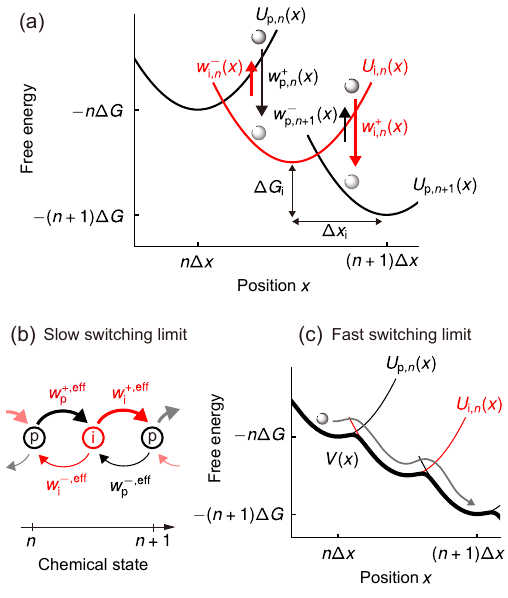}
    \caption{Detailed modeling of the reaction and diffusion and the coarse-graining in the small and large $\wzero$ limit.
    (a) Schematics of the reaction-diffusion model. The black and red curves represent $U_{\mathrm{p}}(x)$ and $U_{\mathrm{i}}(x)$, respectively. The arrows represent the reaction rates between the primary and intermediate states.
    (b) Schematics of the Markov jump processes in the limit of small $\wzero$. Circles represent discrete chemical states, and the arrows represent the effective switching rates among the states.
    (c) Schematics of the effective potential in the limit of large $\wzero$. The black curve shown in bold represents the potential of mean force (PMF) obtained by Eq.~\eqref{eq:PMF}.}
    \label{fig:schematics_kinetic_model}
\end{figure}

We think of a series of states expressed as potential wells, which are alternatively labeled as primary and intermediate states~[Fig.~\ref{fig:intro}a, b].
The states switch by chemical reactions.
The potentials of the primary states are regularly aligned with a distance of $\Delta x$ and assumed to be a harmonic potential with a spring constant $\ko$ for simplicity; $\Uo(x)=\ko(x-n\Delta x)^2/2$.
The potentials of the intermediate states, $\Uint(x)=\kint \left \lbrack x- (n + 1)\Delta x + \Delta \xint \right \rbrack ^2/2$, are located between the potentials of the primary states with a shift of $-\Delta \xint<0$ from the center of the $U_{\mathrm{p}, n+1} (x)$ and have a spring constant $\kint$.
Here, $n$ and p (or i) collectively denote the chemical degree of freedom.
The diffusion in each potential is modeled by the overdamped Langevin equation:
\begin{align} \label{eq:Langevin_eq}
    \gamma \frac{\mathrm{d} x}{\mathrm{d} t} = - \frac{\mathrm{d} U(x)}{\mathrm{d} x} + \sqrt{2 \gamma  \kBT} \xi.
\end{align}
Here, $\gamma$ is a friction coefficient, $\kB$ is the Boltzmann constant, $T$ is the absolute temperature of the surrounding environment, and $\xi$ is the Gaussian white noise with zero mean and unit variance.
$U(x)$ denotes $\Uo(x)$ or $\Uint(x)$ depending on the chemical state.

In the absence of the intermediate states, we consider a forward reaction $(\mathrm{p}, n) \to (\mathrm{p}, n+1)$ and a backward reaction $(\mathrm{p}, n+1) \to (\mathrm{p}, n)$, with rates of $w_{n}^{+} (x)$ and $w_{n+1}^{-} (x)$, respectively~[Fig.~\ref{fig:schematics_kinetic_model}a].
The local detailed balance condition is imposed on the ratio of these rates~\cite{Hill}:
\begin{align}
    \frac{w_{n}^{+} (x)}{w_{n+1}^{-} (x)} = e^{\left \lbrack \Uo(x) - U_{\mathrm{p}, n + 1}(x) + \Delta \Gtot\right \rbrack/ \kBT}.
\end{align}
Here, we denote $ - \Delta \Gtot < 0$ as the free energy change associated with the change in $n$.
There is an arbitrariness in determining each rate under this constraint.
Here, we adopt the following model proposed previously~\cite{KAWAGUCHI20142450}:
\begin{align}\label{eq:rates:wo}
    \begin{split}
    w_{n}^{+} (x) &= \wzero e^{q \left \lbrack \Uo(x) - U_{\mathrm{p}, n+1}(x) + \Delta \Gtot  \right \rbrack/ \kBT}, \\
    w_{n+1}^{-} (x) &= \wzero e^{(q - 1) \left \lbrack \Uo(x) - U_{\mathrm{p}, n+1}(x) + \Delta \Gtot  \right \rbrack/ \kBT},
    \end{split}
\end{align}
where $\wzero$ is the bare switching rate and is, for example, proportional to ATP concentration for ATP-binding events.
The parameter $q\, (0\leq q\leq 1)$ is introduced to specify the asymmetry in the $x$-dependence of the forward and backward switching rates.
This model is simple but can cover a broad range of systems.
Specifically, we can modulate the positional distribution where potential switchings occur by changing $q$.
For example, $q=0$ reproduces the experimental results for the ATP-binding dwells of the ATP hydrolysis by F$_1$-ATPase, where the ATP binding rate does not depend on the shaft angle, but its reverse rate has a significant angle dependence~\cite{KAWAGUCHI20142450, nakayama2024}. 

In the presence of intermediate states, we consider the following four reactions; $(\mathrm{p}, n) \to (\mathrm{i}, n)$, $(\mathrm{i}, n) \to (\mathrm{p}, n+1)$, $(\mathrm{p}, n+1) \to (\mathrm{i}, n)$, and $(\mathrm{i}, n) \to (\mathrm{p}, n)$, with rates of $\woip(x)$, $\wiop(x)$, $\woim(x)$, and $\wiom(x)$, respectively~[Fig.~\ref{fig:schematics_kinetic_model}a].
The local detailed balance conditions are
\begin{align} 
    \frac{\woip}{\wiom} &= e^{\left \lbrack \Uo(x) - \Uint(x) + \Delta \Gtot - \Delta \Gint \right \rbrack/ \kBT}, \label{eq:LDBs} \\
    \frac{\wiop}{\woim} &= e^{\left \lbrack \Uint(x) - U_{\mathrm{p}, n+1}(x) + \Delta \Gint \right \rbrack/ \kBT}.
\end{align}
We denote the chemical free energy change between the $n$-th intermediate state and $(n+1)$-th primary state as $ - \Delta \Gint < 0$.
We use a bare switching rate of $2 \wzero$, since each reaction cycle contains two substeps, in order to legitimately compare the kinetics with and without intermediate states:
\begin{align}\label{eq:rates}
\begin{split}
    \woip (x) &= 2\wzero e^{q \left \lbrack \Uo(x) - \Uint(x) + \Delta \Gtot - \Delta \Gint \right \rbrack/ \kBT}, \\
    \wiom (x) &= 2\wzero e^{(q - 1) \left \lbrack \Uo(x) - \Uint(x) + \Delta \Gtot - \Delta \Gint \right\rbrack/ \kBT}, \\
    \wiop (x) &= 2\wzero e^{q \left \lbrack \Uint(x) - U_{\mathrm{p}, n + 1}(x) + \Delta \Gint \right \rbrack/ \kBT}, \\
    \woim (x) &= 2\wzero e^{(q - 1) \left \lbrack \Uint(x) - U_{\mathrm{p}, n + 1}(x) + \Delta \Gint \right \rbrack/ \kBT}.
    \end{split}
\end{align}
For simplicity, we use the same $q$ values for both $\woip(x)$ and $\wiop(x)$.
The switching rates satisfy translational symmetry; for example, $w_{n + 1}^+ (x) = w_{n}^+ (x - \Delta x)$ and $w_{{\mathrm{p}, n + 1}}^+ (x) = w_{\mathrm{p}, n}^+ (x - \Delta x)$. 
We use the following parameters; $\Delta \Gtot = 10 \, \kBT$, $\Delta x = 6 \, \si{nm}$, $\gamma = 100 \,\kBT\,\si{s/nm^2}$, and $\ko = 2.4 \,\kBT/\si{nm^2}$, assuming a hypothetical molecular motor.
The conclusions of this paper do not qualitatively change with the choice of these parameters.
We numerically calculated the steady-state flux by varying a parameter set of $\Delta\Gint$, $\Delta\xint$, $\kint$, $q$, and $\wzero$.
Specifically, we investigated how these parameters affect the steady-state flux.

\subsection{Flux in Steady State}

We evaluate the steady-state flux in the absence of the intermediate states as
\begin{align}
    J_0 = \int \mathrm{d}x \left \lbrack  w_{n}^{+} (x) \Pbss (x) - w_{n}^{-} (x) \Pbss(x) \right \rbrack,
\end{align}
where $\Pbss (x)$ is the steady-state conditional probability density for the primary states given the $n$-th chemical state, which is normalized as $\int^\infty_{-\infty}\mathrm{d}x\Pbss(x)=1$.
The steady-state probability density satisfies translational symmetry; for example, $P_{\mathrm{p}, n-1}^{\mathrm{ss}} (x,t) = \Pbss (x + \Delta x,t)$.
$\Pbss(x)$ is obtained by solving the Fokker-Planck equation:
\begin{align} \label{eq:Fokker-Planck-wo}
    & \frac{\partial \Pb}{\partial t} =  \left ( \frac{\partial }{\partial x}  \left\lbrack \frac{1}{\gamma} \frac{\mathrm{d} \Uo}{\mathrm{d} x}\right\rbrack + \frac{\kBT}\gamma \frac{\partial^2}{\partial x^2} \right ) \Pb \notag \\
    &  - \left ( w_{n}^{+} + w_{n}^-  \right )\Pb + w_{n+1}^- P_{\mathrm{p}, n+1} + w_{n-1}^+ P_{\mathrm{p}, n-1},
\end{align}
for the steady state by a method previously reported in \cite{nakayama2024}.
The second and subsequent terms on the right-hand side represent the outflow and inflow, respectively.
With the intermediate states, the steady-state flux is calculated as
\begin{align} \label{eq:flux}
    J = \int \mathrm{d}x \left \lbrack  \woip (x) \Pbss (x) - \wiom (x) \Pintss(x) \right \rbrack, 
\end{align}
where $\Pbss (x)$ and $\Pintss(x)$ are the steady-state conditional probability densities for the primary and the intermediate states, respectively, which are normalized as $\int^\infty_{-\infty}\mathrm{d}x[\Pbss(x)+\Pintss(x)]=1$.
$\Pbss(x)$ and $\Pintss(x)$ are obtained as the steady-state solution of the Fokker-Planck equations:
\begin{align}\label{eq:Fokker-Planck}
    &\frac{\partial \Pb}{\partial t}  =  \left ( \frac{\partial }{\partial x}  \left\lbrack \frac{1}{\gamma } \frac{\mathrm{d} \Uo}{\mathrm{d} x}\right\rbrack + \frac{\kBT}\gamma \frac{\partial^2}{\partial x^2} \right ) \Pb \notag \\
    &  - \left ( w_{\mathrm{p}, n}^{-} + \woip  \right )\Pb + w_{\mathrm{i}, n-1}^{+} P_{\mathrm{i}, n-1} + \wiom \Pint,\\
    & \frac{\partial \Pint}{\partial t} 
    =  \left ( \frac{\partial }{\partial x}  \left\lbrack \frac{1}{\gamma} \frac{\mathrm{d} \Uint}{\mathrm{d} x}\right\rbrack + \frac{\kBT}\gamma \frac{\partial^2}{\partial x^2} \right ) \Pint \notag \\
    &  - \left ( \wiom + \wiop \right )\Pint + \woip \Pb + \woim P_{\mathrm{p}, n+1}.
\end{align}

\begin{figure*}[t]
    \centering
    \includegraphics{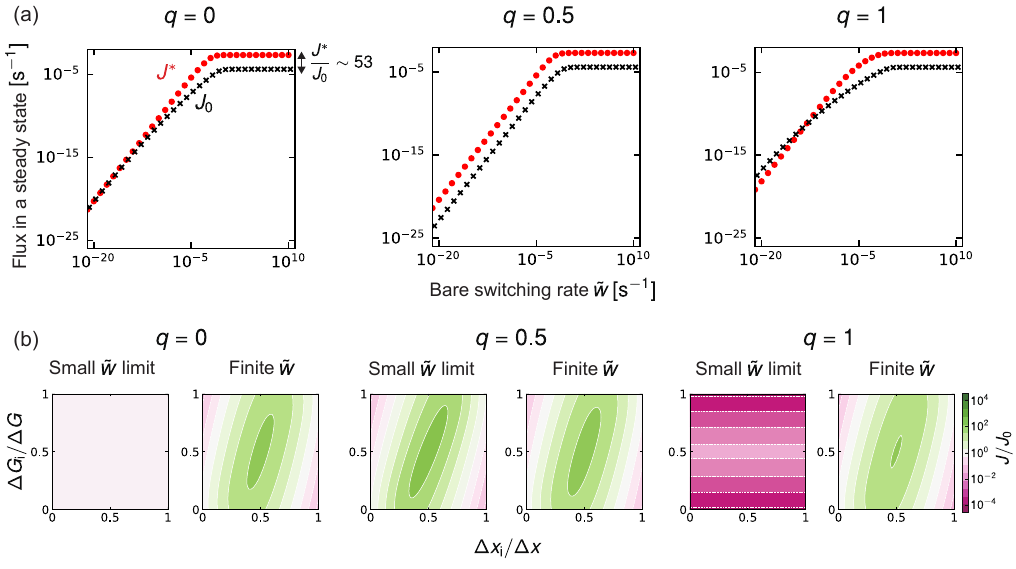}
    \caption{Kinetics with or without intermediate states.
    (a) Dependence of the maximum flux with the intermediate states $\Jmax$ (red) and the flux without intermediate states $J_0$ (black) on the bare switching rate $\wzero$. $\kint = \ko$. 
    (b) Normalized flux $J/J_0$ in the small $\wzero$ limit and for finite $\wzero$ ($\wzero = 10^{-2} \, \si{s^{-1}}$).
    The flux in the small $\wzero$ limit is obtained using Eqs.~\eqref{eq:flux-in-small-w0-limit-wo} and~\eqref{eq:flux-in-small-w0-limit}.
    }
    \label{fig:w0_dependence}
\end{figure*}

In the limit of small $\wzero$, $J$ is analytically obtained.
The distribution of $x$ is equilibrated in each potential because the potential switching is significantly slow.
Therefore, chemical reactions can be described as Markov jumps between the states with effective switching rates~[Fig.~\ref{fig:schematics_kinetic_model}b].
In the absence of the intermediate states, the effective switching rates between $(\mathrm{p}, n)$ and $(\mathrm{p}, n+1)$ are
\begin{align}
    w^{\mathrm{+, eff}} = \int \mathrm{d}x ~w_{n}^{+}(x) P_{\mathrm{p}, n}^{\mathrm{eq}}(x), \\
    w^{\mathrm{-, eff}} = \int \mathrm{d}x ~w_{n}^{-}(x) P_{\mathrm{p}, n}^{\mathrm{eq}}(x),
\end{align}
where $P_{\mathrm{p}, n}^{\mathrm{eq}}(x)$ is the equilibrium probability density conditioned on the $n$-th primary state:
\begin{align}
    P_{\mathrm{p}, n}^{\mathrm{eq}}(x) = \frac{e^{- \Uo(x) / \kBT}}{\int \mathrm{d}x \, e^{- \Uo(x) / \kBT}}.
\end{align}
Then, the steady-state flux is
\begin{align} \label{eq:flux-in-small-w0-limit-wo}
    J_0 = w^{\mathrm{+, eff}} - w^{\mathrm{-, eff}}.
\end{align}
In the presence of the intermediate states, the effective switching rates between $(\mathrm{p}, n)$ and $(\mathrm{i}, n)$ are
\begin{align} \label{eq:forward-rate-small-w0}
     w^{+, \mathrm{eff}}_\mathrm{p} &=  \int  \mathrm{d}x ~\woip(x) P_{\mathrm{p}, n}^{\mathrm{eq}}(x), \\
     w^{-, \mathrm{eff}}_\mathrm{i} &=  \int  \mathrm{d}x ~\wiom(x) P_{\mathrm{i}, n}^{\mathrm{eq}}(x),
\end{align}
where $P_{\mathrm{p}, n}^{\mathrm{eq}}(x)$ and $P_{\mathrm{i}, n}^{\mathrm{eq}}(x)$ is the equilibrium probability densities conditioned on the $n$-th primary and intermediate states.
The forward and backward effective switching rates between $(\mathrm{i}, n)$ and $(\mathrm{p}, n+1)$, denoted by $\wiopeff$ and $\woimeff$, are defined analogously.
Then, the steady-state flux is given as
\begin{align} \label{eq:flux-in-small-w0-limit}
     J = \frac{ w_{\mathrm{p}}^{+, \mathrm{eff}} w_{\mathrm{i}}^{+, \mathrm{eff}} - w_{\mathrm{p}}^{-, \mathrm{eff}} w_{\mathrm{i}}^{-, \mathrm{eff}} }{ w_{\mathrm{p}}^{+, \mathrm{eff}} + w_{\mathrm{i}}^{+, \mathrm{eff}} + w_{\mathrm{p}}^{-, \mathrm{eff}} + w_{\mathrm{i}}^{-, \mathrm{eff}}}.
\end{align}

In the limit of large $\wzero$, the chemical reactions locally equilibrate at each $x$.
In this situation, the dynamics of $x$ obeys Eq.~\eqref{eq:Langevin_eq} with replacing $U(x)$ by the effective potential called the potential of mean force (PMF)~[Fig.~\ref{fig:schematics_kinetic_model}c].
The PMF is given as
\begin{align}
    V_0 (x) = - \kBT \ln\left(\sum_n e^{- \frac{\Uo (x)}{\kBT }} \right), \label{eq:PMF-wo}
\end{align}
and
\begin{align}
    V (x) = - \kBT \ln\left(\sum_n \left[e^{- \frac{\Uo (x)}{\kBT }}+e^{- \frac{\Uint(x)}{\kBT} }\right]\right), \label{eq:PMF}
\end{align}
in the absence and presence of the intermediate states, respectively.
Equations~(\ref{eq:PMF-wo},~\ref{eq:PMF}) are independent of the parameter $q$, that is, the free energy landscape $\Uo (x)$ and $\Uint (x)$ solely determine the PMF, and the asymmetry in kinetics does not affect it.

\section{Results}

\subsection{Effect of intermediate states on the flux}

We investigate the impact of the intermediate states on the kinetics. First, for simplicity, we consider the situation that the primary and intermediate states have the same spring constants; $\ko=\kint$.
In particular, we evaluate the maximum flux with the intermediate states, $\Jmax$, which is obtained by optimizing the positions of the intermediate states specified by $(\Delta \xint, \Delta \Gint)$.
Figure~\ref{fig:w0_dependence}a shows that $\Jmax$ increases with $\wzero$ and saturates in a manner similar to the Michaelis-Menten kinetics~\cite{Michaelis-Menten-original}.
Under the present condition with $\ko=\kint$, $\Jmax$ is obtained when the intermediate state is located at the middle between the primary states; $\Delta\Gint=\Delta\Gtot/2$ and $\Delta\xint=\Delta x/2$.

We found that when $\wzero$ is relatively large, $\Jmax$ is larger than $J_0$ at any given $q$ between 0 and 1, implying the acceleration by the intermediate states.
In the following analysis, the value of $\wzero$ is set to $10^{-2} \,\si{s^{-1}}$ for the finite $\wzero$ case.
With the finite $\wzero$ values, $J$ exceeds $J_0$ except when the intermediate states are located close to the primary states~[Fig.~\ref{fig:w0_dependence}b].
In contrast, in the limit of small $\wzero$, $\Jmax$ is smaller than $J_0$ even with the intermediate states positioned middle when $q$ is close to 0 or 1~[Fig.~\ref{fig:w0_dependence}b; see also Fig.~\ref{fig:dependence-flux-on-q-low-limit} for details].

\begin{figure}[b] 
    \centering
    \includegraphics{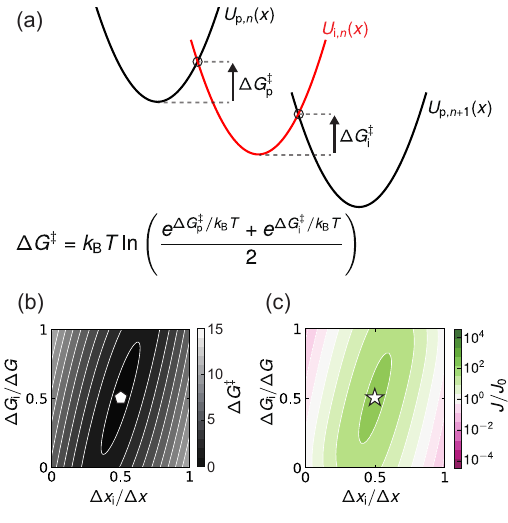}
    \caption{The energetic barriers determine the optimal position for maximizing the flux. (a) The energetic barrier is defined as the height difference between the potential bottom and the intersections of the potentials.
    (b) Effective barrier height given by Eq.~\eqref{eq:effective delta G}. 
    The pentagon indicates the minimum.
    (c) Normalized flux $J / J_0$ for $\wzero = 10^{-2} \,\si{s^{-1}}$ and $q = 0$.
    The star indicates the maximum.}
    \label{fig:energetic_barrier}
\end{figure}

We consider the mechanism of how the intermediate states increase the flux.
In one-dimensional potentials with periodic landscapes, the flux is generally larger when the barrier height between the wells is lower.
As an analogy to this situation, we discuss how the barrier height affects the flux.
We introduce $\Delta G^{\ddagger}_{\mathrm{p}}$ and $\Delta G^{\ddagger}_{\mathrm{i}}$ as the barrier heights for the forward transitions~[Fig.~\ref{fig:energetic_barrier}a].
$\Delta G_{\mathrm{p}}^\ddagger$ is the free energy gap between the local minimum of $\Uo(x)$ and the intersection of the primary and intermediate potentials, $\Uo(x)$ and $\Uint(x)$.
$\Delta G_{\mathrm{i}}^\ddagger$ is defined in the same manner.
Assuming the Arrhenius rates, the average time required for the motor to overcome the barrier is proportional to $e^{\Delta G_{\mathrm{p}}^\ddagger / \kBT}$ or $e^{\Delta G_{\mathrm{i}}^\ddagger / \kBT} $.
Then, the total turnover time for a reaction cycle of $\mathrm{p\to i\to p}$ to complete is approximately proportional to $e^{\Delta G_{\mathrm{p}}^\ddagger / \kBT}+e^{\Delta G_{\mathrm{i}}^\ddagger / \kBT} $.
Hence, we define the effective barrier height as
\begin{align}\label{eq:effective delta G}
    \Delta G^\ddagger \coloneqq \kBT \ln \left (\frac{e^{\Delta G_{\mathrm{p}}^\ddagger / \kBT} + e^{ \Delta G_{\mathrm{i}}^\ddagger / \kBT} }{2} \right ).
\end{align}
The position that minimizes $\Delta G^{\ddagger}$ [Fig.~\ref{fig:energetic_barrier}b, pentagon] and that maximizes the flux~[Fig.~\ref{fig:energetic_barrier}c, star] are similar, implying the significance of $\Delta G^{\ddagger}$ on the flux.
However, the optimal position for maximizing the flux is not solely determined by the potential landscape. 
For example, in the limit of small $\wzero$, the chemomechanical reactions can be described by the Markov jump processes.
In the specific conditions $q = 0$ and 1, the transition rates are independent of $\Delta G^\ddagger$.
Thus, we cannot generally apply the approach based on $\Delta G^\ddagger$ to discrete models.

\subsection{Acceleration by intermediate states is more significant with smaller spring constants}

\begin{figure}[t]
    \centering
    \includegraphics{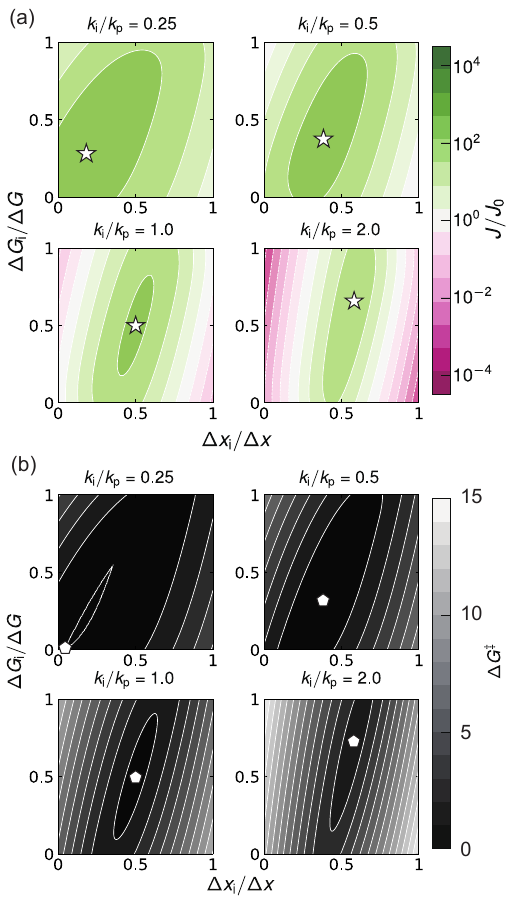}
    \caption{Dependence of normalized flux $J/J_0$ and effective barrier height $\Delta G^\ddagger$ on the position of the intermediate state.
    (a) Heatmap of the normalized flux $J / J_0$  for $\wzero = 10^{-2} \, \si{s^{-1}}$ and $q = 0$.
    The values of $\kint / \ko$ are as indicated.
    The white stars represent the optimal positions for each value of $\kint / \ko$. 
    (b) Heatmap of the barrier height $\Delta G^\ddagger$ with $\kint$.
    The white pentagons represent the positions for minimizing $\Delta G^\ddagger$.}
    \label{fig:optimal_position_high_w0}
\end{figure}

We found that the optimal positions of the intermediate states and the flux ratios $\Jmax/ J_0$ also depend on the spring constant of the intermediate states, $\kint$~[Fig.~\ref{fig:optimal_position_high_w0}a].
Smaller $\kint$ results in larger maximum flux.
As well, $J/J_0$ has a large value in a broader range of $(\Delta \xint, \Delta \Gint)$ for smaller $\kint$, that is, fine tuning is not necessary for the acceleration for small $\kint$.
With large $\kint$, the flux can be smaller than $J_0$ when the position of the intermediate states deviates from the optimal position.

This tendency can be explained in terms of the effective barrier height $\Delta G^{\ddagger}$.
We found a correlation between the position for minimizing $\Delta G^{\ddagger}$~[Fig.~\ref{fig:optimal_position_high_w0}b, pentagon] and the position for maximizing the flux~[Fig.~\ref{fig:optimal_position_high_w0}a, star].
With smaller $\kint$, $\Delta G^{\ddagger}$ becomes smaller, implying a smoother potential landscape.

\begin{figure}[t]
    \centering
    \includegraphics{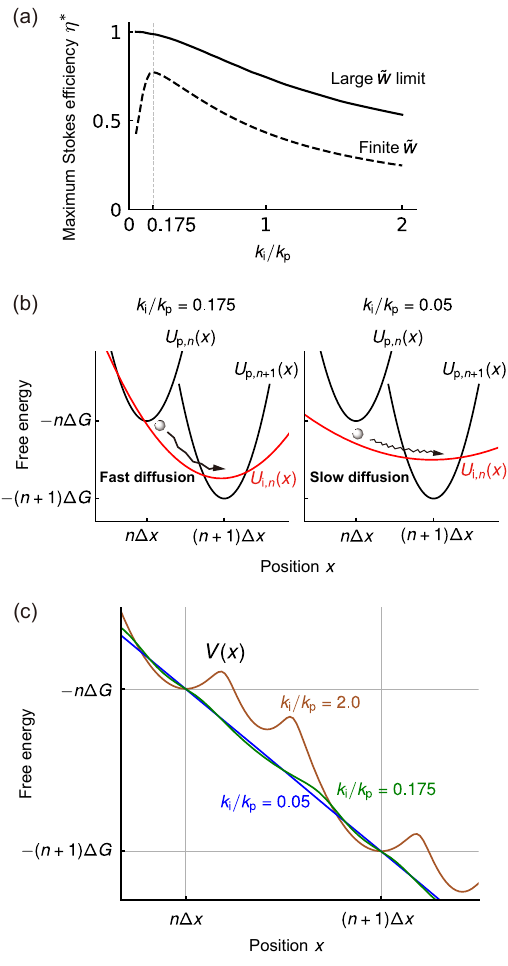}
    \caption{
    Dependence of the maximum Stokes efficiency $\etamax$ and the free energy landscape on $\kint / \ko$.
    (a) $\etamax$ for $\wzero = 10^{-2} \, \si{s^{-1}}$ and large $\wzero$ limit.
    (b) Free energy landscapes with the optimally positioned intermediate states for $\kint / \ko = 0.2$ and 0.05.
    (c) The PMF with the optimal positions of the intermediate states for various $\kint / \ko$ values.
    }
    \label{fig:stokes-eff-vary-ki}
\end{figure}

The smoothness of the landscape can be quantified by the Stokes efficiency~\cite{Hongyun_Wang_200, Li2020}:
\begin{align}
    \eta = \frac{\gamma \left ( J \Delta x \right )^2}{J \Delta \Gtot}. \label{eq:Stokes-efficiency}
\end{align}
$J \Delta \Gtot$ corresponds to the dissipation rate, and $\gamma (J \Delta x)^2$ can be interpreted as the output power against viscous drag.
The maximum Stokes efficiency, denoted as $\etamax$, is obtained for $J=\Jmax$.
We basically observed a decrease in $\etamax$ with $\kint$~[Fig.~\ref{fig:stokes-eff-vary-ki}a], reflecting the fact that $\Jmax$ decreases with $\kint$~[Fig.~\ref{fig:optimal_position_high_w0}a].
The exception is the small $\kint$ region with a finite $\wzero$ value, where a significant decrease in $\etamax$ is observed.
This is due to the slow diffusion on the reduced potential slope~[Fig.\ref{fig:stokes-eff-vary-ki}b].
In the large $\wzero$ limit, $\etamax$ converges to one in the small $\kint$ limit~[Fig.~\ref{fig:stokes-eff-vary-ki}a], where a sigfnificantly smooth PMF is indeed observed~[Fig.~\ref{fig:stokes-eff-vary-ki}c].

\subsection{Acceleration by intermediate states is more significant under external load}

We consider the effect of the intermediate states under external load.
Molecular motors often operate under external load and do positive work in biological cells.
For example, myosin II in skeletal muscle produces positive work when we lift objects~\cite{cell}.
F$_\mathrm{o}$-motor produces mechanical work to rotate F$_1$ for synthesizing ATP molecules~\cite{Okuno-review}.

\begin{figure}[t]
    \centering
    \includegraphics{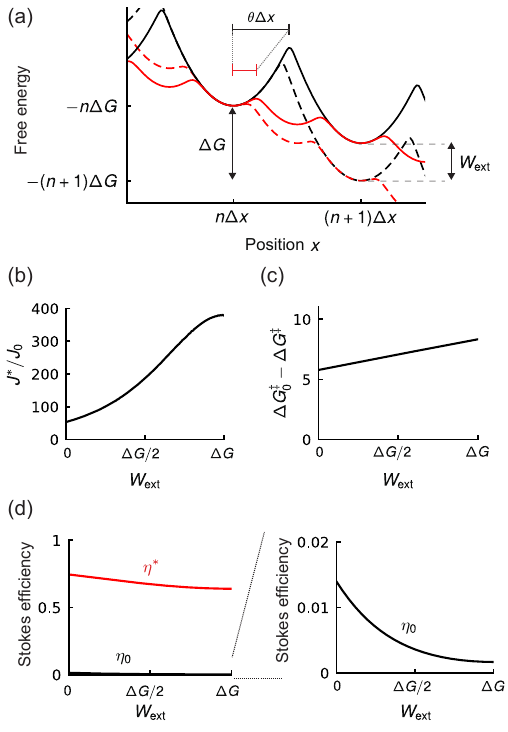}
    \caption{Flux and the Stokes efficiency under external load in the limit of large $\wzero$ and with $\kint = \ko$.
    (a) PMF with and without the intermediate states under an external load. The red and black curves represent the PMF with and without the intermediate states, respectively, in the absence (solid) and presence (dashed) of external load. $\Wext$ is the work against the external load per cycle.
    (b) Dependence of the flux ratio $\Jmax / J_0$ on $\Wext$. 
    (c) Dependence of the difference in barrier heights $\Delta G_0^\ddagger - \Delta G^\ddagger$ on $\Wext$.
    (d) Dependence of the Stokes efficiency $\etamax$ and $\eta_0$ of the motor on $\Wext$.
    }
     \label{fig:max_flux_varied_dissipation}
\end{figure}

We first consider the large $\wzero$ limit, where the potential landscape is approximated by the PMF, for simplicity.
The effect of an external load is considered to be the inclination of the potential landscape, which reduces the overall slope and generally emphasizes the barrier~[Fig.~\ref{fig:max_flux_varied_dissipation}a].
The inclination per $\Delta x$ corresponds to the external work $\Wext$ against the load per cycle. 
The dissipation per cycle decreases to $\Delta \Gtot- \Wext$, which generally reduces flux.
We found that the intermediate states accelerate the flux more significantly in the presence of an external load~[Fig.~\ref{fig:max_flux_varied_dissipation}b], which is consistent with the previous results obtained for Markov jump processes~\cite{Wagoner2019}.
The significant acceleration under load can be explained in terms of the barrier height.
The barrier height for the forward reaction without the intermediate states is denoted as $\Delta G_0^\ddagger$.
We observed the positive correlation between the acceleration~[Fig.~\ref{fig:max_flux_varied_dissipation}b] and the increase in $\Delta G_0^\ddagger - \Delta G^\ddagger$ under an external load~[Fig.~\ref{fig:max_flux_varied_dissipation}c].
In general, the barrier height approximately increases at a rate of $\Wext\theta$ by external load.
Here, $\theta$ is called the load distribution factor~\cite{Howard, Kolomeisky2007} and is defined here as the distance between the potential minimum and the barrier, normalized by $\Delta x$~[Fig.~\ref{fig:max_flux_varied_dissipation}a].
The intermediate states significantly reduce $\theta$ to less than half, as illustrated in Fig.~\ref{fig:max_flux_varied_dissipation}a.
Hence, the external load increases $\Delta G^\ddagger_0$ more significantly than $\Delta G^\ddagger$ and thereby reduces $J_0$ more significantly than $\Jmax$.

\begin{figure}[t]
    \centering
    \includegraphics{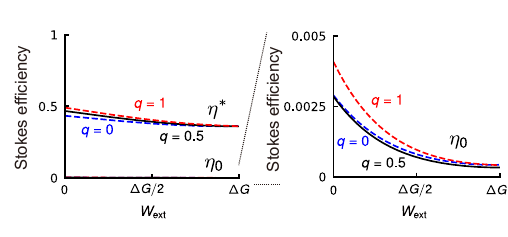}
    \caption{Dependence of maximum Stokes efficiency $\etamax$ and $\eta_0$ on $\Wext$ for various $q$ and $\wzero = 10^{-2} \, \si{s^{-1}}$.
    }
     \label{fig:stokes-eff-finite-w0}
\end{figure}

The effect of the intermediate states under external load is pronounced with the Stokes efficiency.
A large value of $\etamax$ is maintained even under large external load, while $\eta$ in the absence of intermediate states (denoted as $\eta_0$) is significantly small~[Fig.~\ref{fig:max_flux_varied_dissipation}d], consistent with the tendency of $\Jmax/J_0$~[Fig.~\ref{fig:max_flux_varied_dissipation}b].
We observed a similar tendency also in the finite $\wzero$ regime~[Fig.~\ref{fig:stokes-eff-finite-w0}].
These results imply that the molecular motors can leverage the intermediate states to maintain high Stokes efficiency while performing work against an external load.

\subsection{Small $\wzero$ limit}

\begin{figure}[b]
    \centering
    \includegraphics{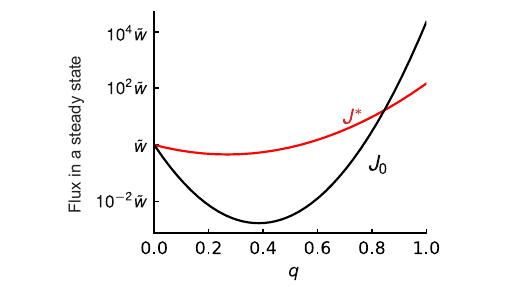}
    \caption{Dependence of the maximum flux $\Jmax$ and $J_0$ on $q$ in the limit of small $\wzero$ for $\kint=\ko$.
    The intermediate states produce less flux than without the intermediate state with large $q$.
    The flux ratios $\Jmax / J_0$ are 0.993 and 0.007 for $q = 0$ and 1, respectively.
    }
    \label{fig:dependence-flux-on-q-low-limit}  
\end{figure}

In the small $\wzero$ regime, we observed the suppression of flux by the intermediate states for certain $q$ values~[Fig.~\ref{fig:w0_dependence}a and \ref{fig:dependence-flux-on-q-low-limit}].
We consider the small limit of $\wzero$, where the kinetics is effectively described by Markov jump process~[Fig.~\ref{fig:schematics_kinetic_model}b], in which the steady-state flux is given by Eqs.~\eqref{eq:flux-in-small-w0-limit-wo} and \eqref{eq:flux-in-small-w0-limit}.
For $q = 0$ and $\kint=\ko$, the effective switching rates,  $\woipeff = \wiopeff =2\wzero$, $\wiomeff = 2\wzero e^{-(\Delta \Gtot - \Delta \Gint )/ \kBT}$, and $\woimeff = 2 \wzero  e^{- \Delta \Gint / \kBT}$, are obtained using Eq.~\eqref{eq:forward-rate-small-w0}.
Then, the steady-state flux is
\begin{align}
    J &= 2 \wzero\frac{ 1-e^{ - \Delta \Gtot / \kBT}}{2 + e^{ - (\Delta \Gtot - \Delta \Gint) \kBT} + e^{ - \Delta \Gint / \kBT}}, \label{eq:J:q=0}\\
    J_0 &= \wzero\left( 1- e^{ - \Delta \Gtot  / \kBT}\right), \label{eq:J0:q=0}
\end{align}
and therefore,
\begin{align}\label{eq:J/J0:q=0}
        \frac J{J_0} &= \frac{1}{1 + \left[e^{-(\Delta \Gtot - \Delta \Gint) / \kBT} + e^{-\Delta \Gint / \kBT}\right]/2}.
\end{align}
Since $J$ is maximized at $\Delta\Gint=\Delta\Gtot/2$,
\begin{align}
\frac{\Jmax}{J_0}=\frac 1{1+e^{ - \Delta \Gtot /2 \kBT}}.
\end{align}
The flux ratio $\Jmax / J_0$ is always less than one, but is close to one with the present parameter $\Delta G=10\,\kBT$.
The increase in $\Delta \Gtot$ reduces the backward rates but does not affect the forward ones.

For $q = 1$, the effective switching rates are $\woipeff = 2\wzero e^{(\Delta \Gtot - \Delta \Gint )/ \kBT}$, $\wiopeff = 2\wzero e^{\Delta \Gint/ \kBT}$, and $\wiomeff =\woimeff = 2 \wzero$, leading to 
\begin{align}
    J &= 2 \wzero\frac{e^{\Delta \Gtot / \kBT} - 1}{2 + e^{(\Delta \Gtot - \Delta \Gint) / \kBT} + e^{\Delta \Gint / \kBT}},\label{eq:J:q=1} \\
    J_0 &= \wzero \left(e^{\Delta \Gtot / \kBT} - 1\right), \label{eq:J0:q=1}
\end{align}
and therefore,
\begin{align}\label{eq:J/J0:q=1}
    \frac J{J_0} &= \frac{1}{1 + \left[e^{(\Delta \Gtot - \Delta \Gint) / \kBT} + e^{\Delta \Gint / \kBT}\right]/2}.
\end{align}
Substituting $\Delta\Gint=\Delta\Gtot/2$ into Eq.~\eqref{eq:J/J0:q=1}, we obtain
\begin{align}
\frac{\Jmax}{J_0}=\frac 1{1+e^{ \Delta \Gtot /2 \kBT}}.
\end{align}
The flux ratio $\Jmax / J_0$ becomes significantly small with $\Delta \Gtot=10\,\kBT$.
The increase in $\Delta\Gtot$ increases the forward rates but does not affect the backward rates.

The dependence of $J/J_0$ on $\Delta\Gint$ in Fig.~\ref{fig:w0_dependence}b is explained by Eqs.~\eqref{eq:J/J0:q=0} and \eqref{eq:J/J0:q=1}.
In general, the flux becomes large when all the forward rates are equal while the backward rates are small~\cite{Geertsema2009}.
For $q$ = 0, the forward rates are constant.
Whereas $\Delta\Gint$ modulates the backward rates, this does not significantly change the flux because the backward rates are typically small, leading to the small sensitivity of $J/J_0$ to $\Delta\Gint$ observed in Fig.~\ref{fig:w0_dependence}b.
For $q=1$, where only the forward rates have $\Delta\Gtot$ and $\Delta\Gint$ dependence, the equal forward rates, $\woipeff=\wiopeff$, are achieved with $\Delta\Gint=\Delta \Gtot / 2$.
Indeed, $J/J_0$ given by Eq.~\eqref{eq:J/J0:q=1} has a steep peak around this optimal position, leading to the high sensitivity to $\Delta\Gint$ observed in Fig.~\ref{fig:w0_dependence}b.
The different sensitivity between $q=0$ and 1 is consistent with the results obtained previously~\cite{Brown2017}.

For $q = 0$ and 1, the flux does not depend on the step size $\Delta x$ nor $\Delta \xint$~ [Eqs.~\eqref{eq:J:q=0}, \eqref{eq:J0:q=0}, \eqref{eq:J:q=1}, and \eqref{eq:J0:q=1} and Fig.~\ref{fig:w0_dependence}b].
This is because the forward or backward rates are constant ($2\wzero$) for $q = 0$ and 1, respectively, and the local detailed balance conditions for the transitions
\begin{align}
    \frac{\woipeff}{\wiomeff} &= e^{(\Delta \Gtot - \Delta \Gint) / \kBT}, \quad
    \frac{\wiopeff}{\woimeff} = e^{\Delta \Gint / \kBT},
\end{align}
also leads to the independence of their reversed rates from $\Delta x$ and $\Delta\xint$.

The dependence of $\Jmax$ and $J_0$ on $q$ in the limit of small $\wzero$ is shown in Fig.~\ref{fig:dependence-flux-on-q-low-limit}.
This dependence is explained as follows.
The forward rate without the intermediate states is
\begin{align}
    w^{\mathrm{+, eff}} &= 2 \wzero e^{ \left \lbrack- \ko \Delta x^2 (1 - q^2 ) / 2 + q \Delta \Gtot \right \rbrack / \kBT}.
\end{align}
On the one hand, the intermediate states split the step size $\Delta x$, which increases the forward rates, leading to larger flux.
$\ko \Delta x^2 (1 - q^2) / 2$ may effectively act as the energetic barrier between the discrete states.
On the other hand, the intermediate states also split $\Delta \Gtot$, reducing the driving force of the motor, leading to smaller flux.
The balance between these two effects determines whether the intermediate states increase the flux or not.
The largest flux is obtained with $q  = 1$ for any given finite $\Delta \Gtot$, which is consistent with the previous results~\cite{Wagoner2016}.

\section{Conclusions}

In this work,  we investigate the effect of intermediate states on the kinetics of molecular motors using a reaction-diffusion model.
The model explicitly considers conformational diffusion in potentials corresponding to chemical states and chemical reactions represented by switching of potentials~[Fig.~\ref{fig:intro}].
The intermediate states accelerate the motor in typical situations~[Fig.~\ref{fig:w0_dependence}].
We observed a correlation between the acceleration and the effective barrier height defined by Eq.~\eqref{eq:effective delta G}, implying a guiding principle that the intermediate states should be positioned to minimize the total barrier heights to maximize flux~[Fig.~\ref{fig:energetic_barrier}].
The acceleration is particularly pronounced in practical situations where a hindering external load is applied to the motor~[Fig.~\ref{fig:max_flux_varied_dissipation}].
This means that the molecular motors can leverage the intermediate states to maintain a large flux while performing work.
However, under specific conditions ($q \simeq 0$ and $q \simeq 1$) in the limit of small $\wzero$, the intermediate states consistently reduce flux even at optimal positions~[Figs.~\ref{fig:w0_dependence} and ~\ref{fig:dependence-flux-on-q-low-limit}].

The present work provides the guiding principle for designing high-performance molecular motors that maximize the flux.
In biology, as illustrated in the introduction using the example of F$_1$, the positioning of the intermediate states can vary even for the same motor from different species.
Detailed modeling of individual motors based on experimentally obtained parameters, such as the spring constants and asymmetric parameters $q$ of each intermediate state, may clarify how the observed step sizes are chosen.

Recently, there has been growing interest in creating artificial molecular motors using synthetic chemistry~\cite{Kassem2017} and DNA nanotechnology~\cite{Bath2007, Wang2019}.
The motor's operating speed is one of the most crucial factors for practical applications (e.g., actuation, transport).
This study may not only contribute to the understanding of biological molecular motors but also help design high-performance artificial molecular motors.

\section*{Acknowledgements}
This work was supported by JSPS KAKENHI Grant Number JP24KJ0403 (to AF), 
JP24K06971 (to YN), JP23H01136 (to ST), and JST ERATO Grant Number JPMJER2302 (to ST).

\end{document}